\def\year{2020}\relax
\documentclass[letterpaper]{article} % DO NOT CHANGE THIS
\usepackage{aaai20-report}  % DO NOT CHANGE THIS
\usepackage{times}  % DO NOT CHANGE THIS
\usepackage{helvet} % DO NOT CHANGE THIS
\usepackage{courier}  % DO NOT CHANGE THIS
\usepackage[hyphens]{url}  % DO NOT CHANGE THIS
\usepackage{graphicx} % DO NOT CHANGE THIS
\usepackage{graphicx}  % DO NOT CHANGE THIS
\usepackage{helvet}  %Required
\usepackage{courier}  %Required
\usepackage[hyphens]{url}
\usepackage{graphicx}  %Required
\usepackage{nicefrac}
\usepackage{booktabs}

\usepackage{algorithm,algorithmic}

\usepackage{amsmath}
\usepackage{amsthm}
\usepackage{mathtools}
\usepackage{amsfonts}
\usepackage{multicol}

		\newcommand{\citet}[1]{\citeauthor{#1}~\shortcite{#1}}
		\newcommand{\citep}{\cite}
\newcommand{\myOmit}[1]{}

\usepackage{color}

\title{Developments in Multi-Agent Fair Allocation}
\author{Haris Aziz\\UNSW Sydney and Data61\\
    Sydney, Australia\\
       haziz@cse.unsw.edu.au}
\begin{document}

\maketitle

\begin{abstract}
Fairness is becoming an increasingly important concern when designing markets, allocation procedures, and computer systems. I survey some recent developments in the field of multi-agent fair allocation. 
\end{abstract}

\section{Introduction}

Many important decisions in our lives are increasingly being made by computers. These decisions include whether we get a certain job or loan. It becomes imperative then that computer systems provide guarantees that the decision has been made transparently and fairly. For many settings such as two-sided matching, fairness considerations based on justified envy or diversity guarantees are enshrined in the letter of the law. Even for smaller day to day decisions, 
the appetite for fairness guarantees is naturally present among people.\footnote{See, for example, the interest in fair allocation portal \url{www.spliddit.org/}).} 
At a bigger scale, the Internet leads to peer to peer interactions where fairness comes up as an important concern~(see, e.g. \citet{Moul18} and \citet{KRT01a}).

The issue of what is fair and how to achieve fairness has been studied in several disciplines including economics~\citep{Moul03a,Thom11b}, political science~\citep{Bram08a,BaYo82a} and philosophy~\citep{Resc02a,Rawl71a}.\footnote{Despite the richness of fairness concepts and principles in these literatures, I will only focus on a handful of fairness concepts in this paper. Other fairness ideas include indices to measure inequality (see, e.g. \citet{Atki70a}) or market-based approaches in which agents have equal budgets (see, e.g. \citet{Vari74a}). } 

I mention some recent developments in fair allocation in multi-agent systems. In particular, I discuss allocation of indivisible items (Section~\ref{sec:indiv}), allocation of divisible items (Section~\ref{sec:div}), two-sided matching (Section~\ref{sec:match}), and fairness in other social choice settings (Section~\ref{sec:sc}). Finally, I wrap up the discussion in Section~\ref{sec:discuss}.

\section{Allocation of Indivisible Items}
\label{sec:indiv}

When indivisible items are allocated among agents, guaranteeing the existence of fair allocations is impossible for many desirable notions of fairness. For example, if there are two agents and one indivisible valued good, allocating the good to either of the two agents will make the other agent envious. In other words, an envy-free allocation may not exist. Another challenge is computational. Suppose agents have positive additive utilities over the items. Then the problem
of checking whether there exists an envy-free allocation is NP-complete even for 1-0 utilities~\citep{AGMW15a}. Similarly the problems of computing allocations that maximize the egalitarian welfare~(utility of the worst off agent) or Nash welfare (geometric mean of the utilities of agents) is NP-hard~(see, e.g. \citet{NNRJ14a}). 
%Does there exists an allocation in which each agent gets at least a certain amount of utility? 

Given these challenges, a couple of influential fairness concepts were proposed by \citet{Budi11a}. The first one is called \emph{maxmin share (MMS)} fairness. The \emph{maxmin fair share} of an agent is the best she can guarantee for herself if she is allowed to partition the items into bundles for the agents but then receives her least preferred bundle. An allocation satisfies MMS fairness if each agent's allocation is at least as preferred to her as her maxmin fair share. Although an MMS fair allocation is not guaranteed to exist even for additive utilities, the concept gives rise to approximation versions in which each agent aspires to get a certain percentage of her maxmin fair share~\citep{PrWa14a}. This has led to several results including new algorithms that are faster, simpler, or provide better approximation guarantees, or work for more general families of utilities~(see e.g. \citet{SGHSY18}, \citet{ARSW17a}, \citet{AMNS15a}, \citet{BaKr17} and \citet{XiLu19a}).

The second solution concept that has been proposed is a relaxation of envy-freeness called \emph{envy-free up to one item (EF1)}. An allocation satisfies EF1 if it is envy-free or any agent's envy for another agent can be removed if some item is ignored. Under additive utilities, 
EF1 can be achieved by a simple algorithm called the round-robin sequential allocation algorithm. Agents take turn in a round-robin manner and pick their most preferred unallocated item. The interest in EF1 was especially piqued when \citet{CKM+16a} proved that for positive additive utilities, a rule based on maximizing Nash social welfare finds an allocation that is both EF1 and Pareto optimal. The result has led to very interesting followup work. For example, \citet{BKV18a} presented a  pseudo-polynomial time algorithm for the same setting to compute an Pareto optimal and EF1 allocation. The returned allocation also provides a 1.45 approximation of the maximum Nash welfare.\footnote{There is a stream of papers primarily focussed on approximating Nash welfare. For example, \citet{CoGz15a} presented a polynomial-time 2.889-approximation algorithm for maximum Nash welfare.}

A majority of the results on indivisible items assume that the items are goods yielding positive utilities (see e.g. \citet{BCM15a,LaRo16}). 
In recent years, there has been a push to obtain similar existence and algorithmic results for negative utilities (see e.g. \citet{Aziz16a}) or  more general utility functions~\citep{PlRo18,OPS19a}. For example, \citet{ACI+18} designed an algorithm that returns in polynomial time an EF1 allocation for doubly monotonic separable preferences.

There are ongoing efforts to develop models and algorithms that take into account real-life features including distributional constraints, dynamic and online settings,  distributed settings, and asymmetric agents. 
An interesting new setting that has been considered is one in which the agents are partitioned into groups and the items are allocated to groups~\citep{SeSu19a}.
Researchers have also been exploring stronger or alternative fairness and efficiency concepts. For example, there is a renewed focus on concepts that concern envy between groups of agents~\citep{AzRe19a,CFS+19,BCEZ19}. Another direction concerns scenarios where items are viewed as nodes of a graph and only those allocations are considered in which each agent gets a connected coalition~\citep{BCE+17a}.  

\section{Allocation of Divisible Items}
\label{sec:div}

When divisible items are allocated among agents, it is easier to guarantee the existence of fair allocations. It can also be relatively easier to compute a fair allocation. Again, consider the example of two agents and one item.
If the item is divisible, we can simply allocate half of the item to each agent, guaranteeing many reasonable fairness criteria including envy-freeness. 

In contrast to the case of indivisible items, an envy-free and Pareto optimal allocation always exists for the case of divisible items. For positive additive utilities, such an allocation can be computed by maximizing Nash social welfare in polynomial time. When the utilities are additive but negative, there are still important gaps in our understanding of the computational complexity of envy-free and Pareto optimal allocations (see e.g. \citet{BMSY17}). 

For divisible items, one important class of utility functions is called \emph{Leontief preferences} in which each agent requires fixed proportions of the items (such as 1 CPU and 5 GB RAM to perform one task). \citet{GZH+11a} proposed a solution for the problem of allocating as much of each item as possible
while equalizing the agents' use of the fraction of their own dominant (bottleneck) resource. The method satisfies envy-freeness, proportionality, Pareto optimality, and strategyproofness. These solutions have  been extended in important ways such as dealing with indivisibilities and dynamic settings~\citep{PPS12a}.

Another model of divisible items is called cake-cutting~\citep{RoWe98a,Proc15a,Proc12a}. A cake is a unit interval representing a heterogeneous item. 
An agent may have different utilities for different subintervals, even if they are of equal size. The challenge is to compute fair allocations using queried information from agents. Cake-cutting has a very long history dating back to the inception of the mathematical theory of fair division. Two of the most well-studied concepts are envy-freeness and proportionality (each agent gets at least $1/n$ of her value of the whole cake). While there are well-known and simple algorithms to compute a proportional allocation, finding envy-free allocations is more challenging. In the last few years, new algorithms have been proposed that compute an envy-free allocation in a number of queries that is bounded in the number of agents. The algorithms work for either positive utilities~\citep{AzMa16c} or negative utilities~\citep{DFHY18a}. 
There is no known bounded algorithm for the case of mixed utilities. 
It has also been noted by the authors that if the requirement to allocate the whole cake is relaxed, then there are simpler and faster algorithms to compute an allocation that is both envy-free and proportional.  In recent years, several new variants of the cake-cutting problems have been studied (see, e.g \citet{SHA15a}).

Recently, researchers have approached problems where indivisible items are allocated among agents, but a minimal number of items are shared or treated as divisible to guarantee existence of fair allocations~\citep{SaSe19b}. 
The approach can be viewed as a natural extension of the Adjusted Winner rule~(see e.g. \citet{BrTo96a}) that finds an envy-free and Pareto optimal allocation among two agents in which only at most one item is shared. 

Money can also be viewed as a divisible item that can be moved around to achieve fairness. A well-studied discrete allocation problem with money is the fair room-rent division problem (see e.g. \citet{ADG91a,ASU04a,GMPZ17a}).

\section{Two-Sided Matching}
\label{sec:match}

In two sided matching markets~(see e.g. \citet{Manl13a}), the goal is to match entities from two different sets. A typical example of the problem is one where 
students/doctors are matched to school/hospitals. Students have preferences over schools. Schools have priorities over students. An outcome in these settings is a matching of students to schools. 

In two-sided matching, an important criterion of desirable outcomes is a fairness property that requires that no student should want to replace another student in a hospital when the hospital has a higher priority for the former student over the latter student. The concept is also based on the idea of envy. The only difference is that an agent can only envy another agent if the former has higher priority.

The property of fairness along with that of non-wastefulness (no student prefers to take some empty slot in a school) are the 
central properties that underpin much of the work on two-sided matching. 
For many general two-sided matching markets, 
the two properties can be simultaneously satisfied by an allocation returned by the well-known Deferred Acceptance algorithm~\citep{Roth08a} or its generalizations. For several important settings such as the allocation of school seats to students~\citep{AbSo03b}, one version of the algorithm additionally satisfies strategyproofness. 

An important research agenda extends this kind of result to other two-sided matching problems with distributional constraints. These constraints included upper quotas on regions of hospitals~(see e.g. \citet{KaKo15,BFIM10,GIKY+16a}) as well as target lower quotas for diverse representation~\citep{AGSW19a,EHYY14a,KHIY15a}.  These new settings give rise to fundamental questions of tradeoffs between merit, diversity, and segregation. Yet another direction is taking a time-sharing approach to two-sided matching (see e.g. \citet{AzKl19b,KeUn15a}).

\section{Fairness in other Social Choice Settings}
\label{sec:sc}

%The domain of fair allocation is 
We have mostly focussed on allocations problems where agents get ownerships of their allocation. 
We can also consider allocation of public goods such as the funding of public projects. Public goods can be selected via voting on the different options. 
In classical voting settings in which a single alternative is to be selected, it is very difficult to guarantee reasonable representation of sizable minorities. Consider the case where 51$\%$ of the agents most prefer alternative $a$ and 49$\%$ of the agents prefer alternative $b$. Then alternative $a$ that is supported by a majority is selected even though the remaining agents may find it unpalatable. 

There are other voting settings where fairness concerns become more meaningful. One such setting is a simple extension of single-winner voting to selecting multiple winners (see e.g. \citet{ABC+16a,FSST17a,Aziz19b}). Another setting is probabilistic voting or portioning in which each alternative is given a certain probability weight or a fraction of a budget (see e.g. \citet{AAC+19a,ABM19a,Aziz18a,Bran17a}). In both settings, meaningful fairness axioms have been proposed and algorithms have been designed to satisfy the axioms.
A general principle while formalizing the axioms is that of proportional representation: if a large enough group cohesively prefers some outcome(s), then that group should get suitable representation (see e.g. \citet{AzLe19a}).  
Recently, fairness concerns are also being examined in participatory budgeting problems in which projects have costs and there is a constraint on the maximum budget that can be used~\citep{FGM16a,ALT18a,FST+17a}. 

Another collective decision-making setting where fairness becomes more meaningful is the case where there are several issues with each issue having its own candidates outcomes~\citep{CF017}. Agents have preferences over outcomes over each of the issues. The setting is more general than allocation of indivisible items in following way. Each issue corresponds to the decision of allocation of a particular item. The possible outcomes of an issue correspond to giving the item to one of the agents. It turns out that a solution based on maximizing Nash social welfare satisfies both Pareto optimality and a natural relaxation of proportionality. 

\section{Discussion}
\label{sec:discuss}

%I have provided a bird's eye view of multi-agent settings in which fairness is an important issue. 
I discussed some settings where fairness plays an important role. Fairness concerns can come up in many other multi-agent settings as well such as auctions and coalition formation. Much of the work in multi-agent fair allocation is centered around formal axioms to capture fairness.\footnote{In addition to standard axioms such as envy-freeness, a formal theory of fair allocation covers many different axiom for desirable rules or algorithms~(see, e.g. \citet{Thom15a}.)} It will be interesting to use data-centric approaches to understand how humans conceptualize fairness or decide on tradeoffs between societal outcomes. 

Another area of computer science where fairness is taking center stage is machine learning. Machine learning algorithms learn from data to make actionable suggestions such as giving a loan. Whether these algorithm are biased is a prominent concern for society~(see, e.g. \citet{KLR18a}). A natural approach to incorporate fairness in these algorithms follows the Aristotelian ``equal treatment of equals'' principle.
It is an interesting challenge to explore ways in which ideas from multi-agent fair allocation can contribute to fairness in machine learning. In general, the interplay of fairness with other desirable aspects such as accountability, transparency, and trust is broad research question~(see, e.g. \citet{ShPa19a}). 

As computers start to make important decisions in our lives, it is timely to revisit the lessons learnt. New computer applications will give rise to novel and challenging problems so it is critical to continue updating our toolkit of axioms and methods for designing and enabling fair multi-agent systems.

\section*{Acknowledgements}

The survey is based on a presentation given by the author at the Dagstuhl-Seminar on Application-Oriented Computational Social Choice (Seminar Number 19381). The author thanks Alex Lam and the reviewers for useful comments. 

% \clearpage
\bibliographystyle{aaai}
%\bibliographystyle{ACM-Reference-Format}
 % \bibliography{abb,group,aziz,fair,GEF,matching,adt}
 %

\end{document}